\documentclass[twocolumn,showpacs,a4,prl]{revtex4}
\usepackage{graphicx}%
\usepackage{dcolumn}
\usepackage{bm}
\makeatletter
\def\btt#1{\texttt{\@backslashchar#1}}%
\DeclareRobustCommand\bblash{\btt{\@backslashchar}}%
\makeatother

\begin{document}

\preprint{KNIGHTe.TEX}

\title{$^{11}$B and $^{27}$Al NMR spin-lattice relaxation and Knight shift study of 
Mg$_{1-x}$Al$_x$B$_2$. Evidence for anisotropic Fermi surface.}

\author{G. Papavassiliou$^1$, M. Pissas$^1$, M. Karayanni$^1$, M. Fardis$^1$, S. Koutandos$^1$, and K. Prassides$^{1,2}$}
\affiliation{$^1$Institute of Materials Science, NCSR,  Demokritos, 153 10 Aghia Paraskevi, Athens, Greece\\
$^2$School of Chemistry, Physics and Environmental Science, University of Sussex, Brighton BN1 9QJ, UK} 
\date{\today }

\begin{abstract}
We report a detailed study of $^{11}$B and $^{27}$Al NMR spin-lattice 
relaxation rates ($1/T_1$), as well as of $^{27}$Al Knight shift (K) of 
Mg$_{1-x}$Al$_x$B$_2$, $0\leq x\leq 1$. The obtained ($1/T_1T$) and K vs. $x$ plots are
in excellent agreement with ab initio calculations. 
This asserts experimentally the prediction that the Fermi surface is highly anisotropic,
consisting mainly of hole-type $2-D$ cylindrical sheets from bonding $2p_{x,y}$ boron 
orbitals.
It is also shown that the density of states at the Fermi level decreases sharply on Al 
doping and the $2-D$ sheets collapse at $x\approx 0.55$, where the superconductive phase disappears.
\end{abstract}
\maketitle

The discovery of superconductivity in MgB$_2$  attained recently 
a lot of interest \cite{Akamitsu01}, as this binary alloy 
reveals a remarkably high $T_c\approx 40$K. MgB$_2$ is isostructural and isoelectronic with 
intercalated graphite (ICG), with carbon replaced by boron, and therefore exhibits similar bonding and 
electronic properties as ICG. Thus the high $T_c$ value of  MgB$_2$ in comparison to ICG ($\sim 5$K) was very surprising. 

Band structure calculations \cite{Cortus01,An01,Antropov01,Kong01,Singh01} have shown that Mg is substantially ionized in this compound, however the 
electrons donated to the system are not localized on the B anion, but rather are distributed over 
the whole crystal. The six B $p$ bands contribute mainly at the Fermi level. 
The unique feature of MgB$_2$ is the incomplete filling of the two $\sigma$ 
bands corresponding to prominently covalent $sp^2$-hybridized bonding within the graphite-like boron layer.
Two isotropic $\pi$ bands are derived from B $p_z$ states and 
four two dimensional $\sigma$-bands from B $p_{x,y}$. Both $p_z$ bands cross the Fermi level, while only two bonding $p_{x,y}$ bands
and only near the $\Gamma$ point (0,0,0) do so, forming cylindrical Fermi surfaces around $\Gamma$-A line. Due to their
2D character, these bands contribute more than 30\% to the total density of states (DOS)\cite{Cortus01,Antropov01,Pavarini01,Belashchenko01}. Such strong anisotropy in the Fermi surface (and possibly in the electron phonon coupling) conciles with the recently reported anisotropy in $H_{c2}$ \cite{Lima01,Patnaik01,Simon01,Budko01,Papavassiliou02,Pissas02}, and the existence of two superconducting gaps \cite{Bouqet01,Liu01,Szabo01,Chen01,Tsuda01,Giubileo01,Yang01}. 

\begin{figure}[tbp] \centering
\includegraphics[angle=0,width=8.5cm]{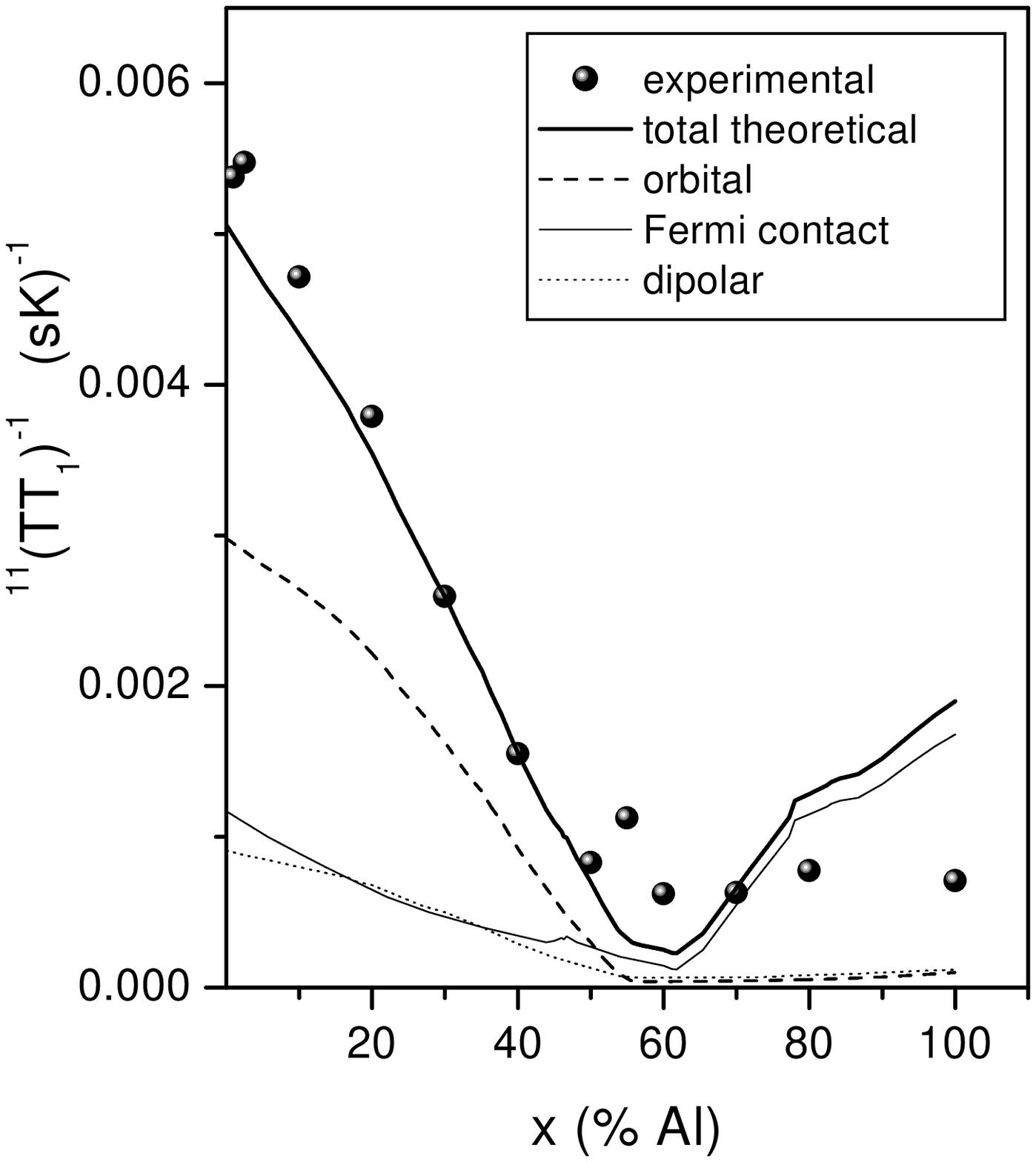}
\caption{Boron $^{11}$(1/T$_1$T) for Mg$_{1-x}$Al$_x$B$_2$ as a function of Al-doping. Lines show the ab initio calculated plots from Refs. \onlinecite{Antropov01,Belashchenko01}}\label{fig1}
\end{figure}

In view of these important findings, great interest has been raised on measurements of electron or hole doped MgB$_2$, aiming to clarify how the electron DOS and the Fermi surface depend on doping. 
A very suitable substitution for such a study is Al, which donates three electrons 
(instead of two for Mg), and thus  doping by one electron per atom. 
In addition, the end members MgB$_2$ and AlB$_2$
as well as the intermediate mixed crystals Mg$_{1-x}$Al$_x$B$_2$ crystallize in the $P6/mmm$ space group, whereas  by Al doping the lattice constants decrease almost linearly \cite{Slusky01}. The similarity of the calculated electronic density of states between MgB$_2$ and AlB$_2$ indicates that Al doping results in simple filling of the available electronic states. Suzuki {\it et al.} \cite{Suzuki01} predicted that in Mg$_{1-x}$Al$_x$B$_2$ the concentration of $\sigma$ holes varies with $x$ as $n_h=(0.8-1.4x)\times 10^{22}$ cm$^{-3}$, leading to $n_h=0$ for $x\approx 0.6$. 
A similar conclusion was deduced in Refs. \onlinecite{An01,Antropov01,Pena02}. Under these aspects, the detrimental effect of Al doping on $T_c$ can be explained by the fact that doping increases the Fermi energy ($E_F$), while decreasing the DOS $N(E_F)$. 

An excellent probe to study the influence of the Al substitution on the electronic structure of electron doped 
MgB$_2$ is nuclear magnetic resonance (NMR). Knight shift, $K$, and nuclear spin-lattice 
relaxation (NSLR) rate $1/T_1$ measurements, give us the possibility to determine experimentally 
$N(E_F)$ through the static and fluctuating parts of the hyperfine field, induced at 
the position of the resonating nuclei from electrons at the Fermi surface. This allows to estimate the contribution of 
different atoms to $N(E_F)$ as well as and the anisotropy of electronic states at the Fermi level. 

The Hamiltonian describing the magnetic interaction of the nucleus with the atomic electrons can be written as \cite{Abragam}:
${\cal H}=2(8\pi/3)\mu_B\gamma_n\hbar {\bf I}\cdot{\bf S(r)}\delta({\bf r})
-2\mu_B\gamma_n\hbar {\bf I}\cdot[{\bf S}/r^3-3{\bf r}({\bf S}\cdot {\bf r})/r^5]
-\gamma_n\hbar (e/mc)[{\bf I}\cdot({\bf r\times p})/r^3]$, 
where $\mu_B$ is the Bohr magneton, $\gamma_n$ the gyromagnetic ratio, ${\bf I}$ and ${\bf S}$ the nuclear and 
electron spins respectively, and ${\bf r}$ is the radius vector of the electron with the nucleus at the origin.  
In the formula above, the first term describes the Fermi contact interaction,  the second term the spin dipolar interaction between nuclear and electron spins, and the third term the coupling with the electronic orbital moment. In the simplest case, where only contribution from the Fermi contact term is considered, the first term can be rewritten as 
${\cal H}_{KS}\propto -V(8\pi/3)\gamma_n\hbar \chi_p\langle |\Psi(0)|^2\rangle_{F_S}{\bf I\cdot H_0}$, 
where $\chi_p=M/H=\mu_B^2 N_s(E_F)/V$, and the symbol $\langle\  \rangle_{F_S}$
means the average over all s orbitals at the Fermi surface. 
Due to this term the nuclear spin ${\bf I}$ 'sees' an internal field 
$V\chi_p{\bf H_0}/2\mu_B$, which is superimposed on the applied external magnetic field, 
and causes a paramagnetic shift of the nuclear resonance, i.e. the Knight shift. 
In a similar way, the relaxation rate from the Fermi contact term is expressed by the 
relation \cite{Abragam}:
$(1/T_1)=(64\pi^3/9)\gamma_e^2\gamma_n^2\hbar^3\langle |\Psi_{\bf k}(0)|^2|\Psi_{\bf k'}(0)|^2\rangle_F N_s(E_F)^2k_BT$. A similar dependence on N$_{2p}$(E$_F$)$^2$ holds if the nuclear Hamiltonian is dominated by the nuclear-electron orbital interaction \cite{Obata63}. Hence, the dependence of both K and $1/T_1$ on certain partial $N_i(E_F)$ is clear.

\begin{figure}[tbp] \centering
\includegraphics[angle=0,width=8.5cm]{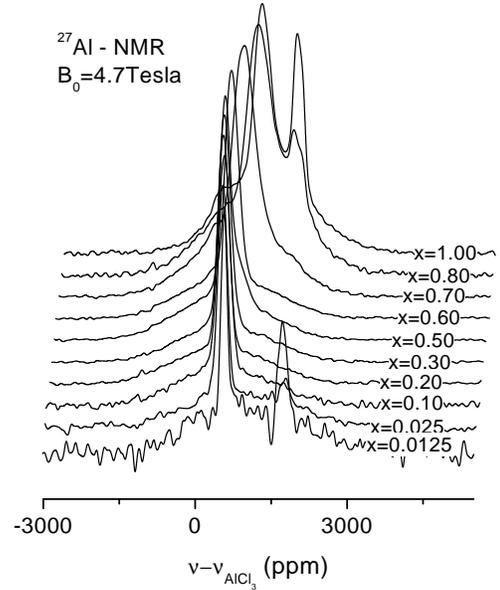}
\caption{$^{27}$Al NMR line shapes of the central transition at room temperature for Mg$_{1-x}$Al$_x$B$_2$. For $x=0.0125$ and $0.025$ $100,000$ accumulations were acquired, due to the weakness of the signals. In the low doping regime, the spurious signal at $+1700$ ppm is coming from the probe. The peak at the same frequency for $x=0.80$, $1.0$ is produced by free Al that unavoidably remained during sample preparation. For clarity, all spectra are normalized to one.}\label{fig2}%
\end{figure}

Until now, $^{11}$B, $^{27}$Al and $^{25}$Mg NSLR and Knight shift measurements have been reported for pure MgB$_2$ and AlB$_2$ \cite{Kotegawa01,Jung01,Baek02,Mali02,Gerashenko02}, which are in agreement with the theoretical predictions. These results, in conjunction with ab initio calculations \cite{Antropov01,Pavarini01,Belashchenko01} have shown that in MgB$_2$ the $^{11}$B NSLR is dominated by orbital relaxation, whereas in AlB$_2$ the $^{11}$B NSLR is overruled by the Fermi-contact interaction. 
On the other hand, $^{27}$Al and $^{25}$Mg NSLRs, as well as the Knight shift on all three $^{11}$B, $^{27}$Al, and $^{25}$Mg sites was shown to be controled by the Fermi-contact polarization \cite{Pavarini01,Baek02}. Besides, $^{11}$B-NMR NSLR relaxation rate measurements on mixed Mg$_{1-x}$Al$_x$B$_2$, $x\leq 0.2$, have shown a rapid decrease of 1/(T$_1$T) with doping that was attributed to reduction of the total N(E$_F$) \cite{Kotegawa02}. Nevertheless, a detailed NMR study of the variation of each partial N$_i$(E$_F$) with Al doping, and  comparison with theory is lacking so far. 

\begin{figure}[tbp] \centering
\includegraphics[angle=0,width=8.5cm]{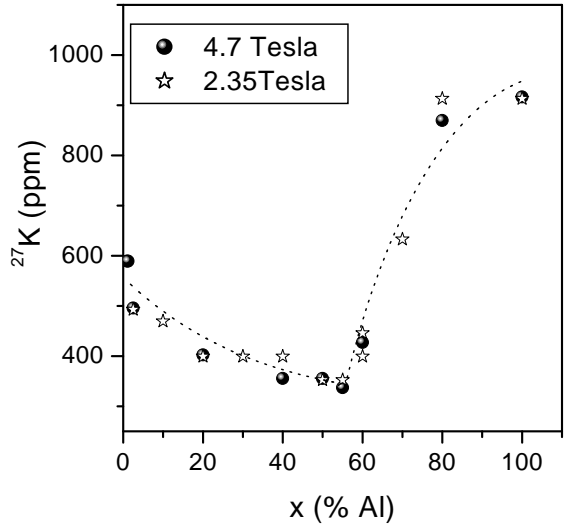}
\caption{The $^{27}$K Knight shift of  NMR spectra for Mg$_{1-x}$Al$_x$B$_2$ in fields $2.35$ and $4.7$ Tesla.}\label{fig3}%
\end{figure}

The purpose of this work is to report a systematic study of $^{11}$B and $^{27}$Al NSLR rates 1/T$_1$, as well as of $^{27}$Al Knight shifts, as a function of Al doping for Mg$_{1-x}$Al$_x$B$_2$, $0 \leq x \leq 1$. $^{11}$B Knight shift measurements were not considered, because the isotropic $^{11}$B Knight shift is small ($+40$ ppm for MgB$_2$ and $-10$ ppm for AlB$_2$ \cite{Baek02}), and of the same order of magnitude with the dipolar and the second order quadrupolar split in the NMR fields $2.35$, and $4.7$ Tesla that have been used in this work. Our measurements show an excellent agreement between the experimental boron $^{11}$($1/T_1T$) plot with that obtained from local density-functional methods \cite{Antropov01,Belashchenko01}, and the dominance of the orbital relaxation up to $x\approx 0.55$, where T$_c$(x) vanishes. This is a convincing evidence that up to this doping the hole-type $2-D$ cylindrical sheets (from bonding 2p$_{x,y}$ boron orbitals) of the Fermi surface play an essential role in the $^{11}$B NSLR. The slight decrease of both $^{27}$K and $^{27}$(1/T$_1$T) for $x\leq 0.55$, and their abrupt increase above this doping value are in support of this conclusion. 

Polycrystalline samples of nominal composition Mg$_{1-x}$Al$_x$B$_2$ for $0\leq x\leq 1$ were prepared by 
reaction of Al and Mg powders with amorphous B at temperatures between 700$^{\rm o}$C and 910$^{\rm o}$C 
as described elsewhere \cite{Pissas02}. We notify that: According to the available literature \cite{Slusky01,Xiang02,Li02,Zanbergen02} and our data, the temperature where the reaction (preparation temperature) takes place, defines the shape of the $(00l)$ diffraction peaks in the region $0.05\leq x\leq 0.5$. The existence of significant broadening or/and splitting in these peaks manifest the existence of some kind of phase separation in this doping range. (ii) Carefully preparated samples in the region arround $x=0.5$ display a broad superlattice peak $(0 0 1/2)$ which means that ordering of Mg and Al occurs \cite{Margadonna02}.

$^{27}$Al NMR line shape measurements of the central transition $(-1/2 \rightarrow 1/2)$ were performed on two spectrometers operating in external magnetic fields $H_0=2.35$ and $4.7$ Tesla. Spectra were obtained from the Fourier transform of half of the echo, following a typical $\pi /2- \tau- \pi /2$ solid spin-echo pulse sequence. 
The $^{11}$B $T_1$ of the central line was determined by applying a saturation recovery technique, and fitting with the two exponential relaxation function that is appropriate for $I=3/2$ nuclei \cite{Andrew61}. Correspondingly, $^{27}$Al $T_1$ was determined by applying the three-exponential recovery law that is appropriate for $I=5/2$ nuclei \cite{Mali02,Andrew61}. 

Figure \ref{fig1} shows the boron $^{11}$(1/TT$_1$) as a function of $x$, in the normal state. In all cases a single component of $T_1$ was found to fit satisfactorily the magnetization recovery curves. Besides, a $1/T_1T=$constant relation was fiting the experimental data from room temperature down to $80$K.  Our measurements reveale that by increasing doping, $^{11}$(1/TT$_1$) decreases rapidly up to $x=0.55$, and subsequently exhibits a slight increase for $x\geq 0.55$. For reasons of comparison, we have plotted the calculated $^{11}$(1/TT$_1$) values from 
Ref. \onlinecite{Belashchenko01}, for all three orbital, dipole-dipole, and Fermi contact term contributions. Clearly, the orbital term dominates in the $^{11}$B relaxation rates for $x\leq 0.55$. In case of pure MgB$_2$ the $^{11}$B orbital hyperfine interaction of $2p$-holes with the nuclear magnetic moments is about 3 times larger than the dipole-dipole, and the Fermi contact interaction. This is due to the fact that the boron p$_\sigma$ and p$_\pi$ bands are all at the Fermi level (N$_{px}$=N$_{py}\approx 0.035$, N$_{pz}\approx 0.045$ states/eV/spin/B), whereas only a few s boron electrons are close to the Fermi level (N$_s\approx 0.002$ states/eV/B)\cite{Pavarini01,Baek02}. This gives a ratio between the Fermi-contact and the orbital/dipole-dipole coupling constants, $F\simeq 0.35$ \cite{Pavarini01}, and $^{11}$T$_1^{-1}$ is mainly proportional to N$_{2p}$(E$_F$)$^2$. It may thus be inferred that the rapid decrease of the $^{11}$B relaxation rate is due to decrease of the DOS in the $2-D$ hole-type sheets, while a minimum 1/T$_1$ value is obtained at $0.55\leq x\leq 0.60$, where the $2-D$ sheets appear to collapse. We also notice the discrepancy between the theoretical and experimental values for $x>0.6$. Most probably, calculations tend to overestimate the Fermi contact interaction at the position of the B nucleus in this doping range \cite{Baek02}.   

\begin{figure}[tbp] \centering
\includegraphics[angle=0,width=8.5cm]{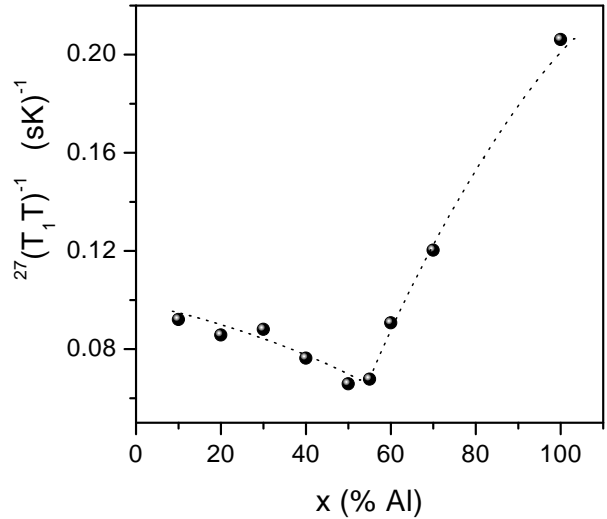}
\caption{$^{27}$(1/T$_1$T) for Mg$_{1-x}$Al$_x$B$_2$ as a function of Al-doping x.}\label{fig4}
\end{figure}

Figure \ref{fig2} demonstrates $^{27}$Al NMR line shapes of Mg$_{1-x}$Al$_x$B$_2$ at room temperature in field $4.7$ Tesla. A completely similar picture was obtained in field $2.35$ Tesla. The spectra for $x=0.0125$, and $0.025$, were extremely weak, and therefore were acquired with $100,000$ signal accumulations (for comparison, signals for $x\geq 0.1$ were acquired with  $512$ accumulations). In all samples the spectra consist of a central transition line, $\approx 20$ kHz wide, which shifts with doping,  and a broad powder pattern from the satellite transitions. The low doping spectra exhibit a spurious weak peak at $+1700$ ppm that was produced by the probe. The strong peak at the same frequency for $x=0.80, 1.0$ is produced by free Al that inavoidably remains during sample preparation at high doping concentrations. 

In Figure \ref{fig3} we show the shift of the central line peak as a function of $x$, in fields $2.35$ and $4.7$ Tesla. The signal of a standard aqueous solution of AlCl$_3$ was used as reference. The coincidence of the curves in both fields is a clear evidence that the obtained spectral shift corresponds solely to the $^{27}$K shift. In typical metallic shifts of $I=5/2$ nuclei like $^{27}$Al, in addition to the Knight shift, the center-of-gravity position of the NMR signal should include the second-order quadrupole shift given by $\Delta \nu =(25\nu _Q^2)/(18\nu _L)$ \cite{Cohen57}. However, recent experiments on AlB$_2$ have shown that the quadrupolar coupling constant is $\nu _Q\simeq 80$ kHz \cite{Baek02}, thus giving a negligibly small second order quadrupolar shift, $\Delta \nu \approx 14$ ppm. According to Figure \ref{fig3}, by increasing $x$ the $^{27}$K decreases rapidly, whereas  for $x\geq 0.55$, i.e. at the doping value where the superconductive phase disappears \cite{Pena02}, it increases sharply becoming $\approx +900$ ppm for pure AlB$_2$. As previously shown, by Al doping of MgB$_2$ the $\sigma$ hole bands are filled \cite{An01,Antropov01,Pena02}, and their contribution to N(E$_F$) becomes zero at $x\approx 0.55$ \cite{Pena02}. Evidently, the gradual decrease of $^{27}$K for $x\leq 0.55$ reflects, (i) the initial slight decrease of $N_s(E_F)$ in this doping range \cite{Cortus01}, and (ii) the reduction of the Stoner enhancement by filling the $\sigma$ hole bands, due to decrease of the total N(E$_F$). We notice that the Stoner enhancement renormalizes both K and $1/T_1$  by a factor $S=1/(1-IN(E_F))^\alpha $, with $\alpha =1$ for K, and $1\leq \alpha \leq 2$ for $1/T_1$ \cite{Pavarini01,Belashchenko01}. On the other hand, the sharp increase of the Knight shift for $x\geq 0.55$, may be attributed to the rapid increase of N$_s$ by further doping, after completely filling the $2p_{x,y}$ hole bands (N$_s$(Mg) in MgB$_2$ is $\approx 0.0092$ states/eV/spin, whereas N$_s$(Al) in AlB$_2$ is $\approx 0.0362$ states/eV/spin \cite{Pavarini01}).  
A similar behaviour is observed in Figure \ref{fig4}, which exhibits the $^{27}$(1/TT$_1$) vs. x plot. This is expected as  the $^{27}$Al relaxation rate is dominated by the Fermi contact interaction, and therefore is proportional to N$_s^2$(Al) \cite{Baek02}. 

In conclusion, $^{11}$B and $^{27}$Al NMR NSLR rate and Knight shift measurements have been employed in order to investigate the structure and the variation of the Fermi surface in MgB$_2$ upon Al (i.e. electron) doping. Our results are completely consistent with calculations predicting a strongly anisotropic Fermi surface that is comprised from hole-type $\sigma$-bonding $2-D$ cylindrical sheets, and a hole-type and electron-type, $3-D$ $\pi$-bonding tubular network. The collapse of the $2-D$ sheets at $x\simeq 0.55$, as predicted by theory, is experimentally verified by the fast decrease of the $^{11}$B NSLR rate for $x\leq 0.55$ and the sharp increase of both the $^{27}$K, and  $^{27}$NSLR rates for $x\geq 0.55$. The latter  indicates a strong reshaping of the Fermi surface towards the electronic structure of AlB$_2$, due to interplane electron contribution. Our results concile with both experimental and theoretical evidence that indicate anisotropic pairing and multi-gap superconductivity in MgB$_2$.

\end{document}